\documentclass[twocolumn,showpacs,superscriptaddress,amsmath,amssymb,aps,pra]{revtex4-1}
\usepackage{bm}
\usepackage{graphicx}
\usepackage{dcolumn}
\usepackage{amsmath,txfonts}
\usepackage{epstopdf}
\usepackage[mathlines]{lineno}
\usepackage{color}
\usepackage[colorlinks=true,linkcolor=blue,citecolor=blue]{hyperref}
\usepackage[normalem]{ulem}

\begin{document}

\title{Generation of long-lived $W$ states via reservoir engineering in dissipatively coupled systems}

\author{Guo-Qiang Zhang}
\affiliation{School of Physics, Hangzhou Normal University, Hangzhou, Zhejiang 311121, China}

\author{Wei Feng}
\affiliation{School of Physics, Hangzhou Normal University, Hangzhou, Zhejiang 311121, China}

\author{Wei Xiong}
\affiliation{Department of Physics, Wenzhou University, Zhejiang 325035, China}

\author{Qi-Ping Su}
\affiliation{School of Physics, Hangzhou Normal University, Hangzhou, Zhejiang 311121, China}

\author{Chui-Ping Yang}
\email{yangcp@hznu.edu.cn}
\affiliation{School of Physics, Hangzhou Normal University, Hangzhou, Zhejiang 311121, China}
\affiliation{Quantum Information Research Center, Shangrao Normal University, Shangrao 334001, China}

\begin{abstract}
Very recently, dissipative coupling was discovered, which develops and broadens methods for controlling and utilizing light-matter interactions. Here, we propose a scheme to generate the tripartite $W$ state in a dissipatively coupled system, where one qubit and two resonators simultaneously interact with a common reservoir. With appropriate parameters, we find the $W$ state is a dark state of the system. By driving the qubit, the dissipatively coupled system will evolve from the ground state to the tripartite $W$ state. Because the initial state is the ground state of the system and no measurement is required, our scheme is easy to implement in experiments. Moreover, the $W$ state decouples from the common reservoir and thus has a very long lifetime. This scheme is applicable to a wide class of dissipatively coupled systems, and we specifically illustrate how to prepare the $W$ state in a hybrid qubit-photon-magnon system by using this scheme.
\end{abstract}

\date{\today}

\maketitle

\section{Introduction}

In the last decades, the coherent coupling between light and matter (light and light, or matter and matter) has been widely studied in various physical platforms due to its diverse applications in quantum communications~\cite{Reiserer15}, quantum computing~\cite{Kurizki15,Gu17}, quantum sensing~\cite{Degen17}, and so on. Not only that, the light-matter coherent coupling is of fundamental importance. For example, the coherent coupling has entered the ultrastrong-coupling regime~\cite{FornDiaz19}, where the counter-rotating terms produce unexpected physical phenomena, such as the virtual photon population in the ground state~\cite{Liberato09} and the Bloch-Siegert shift~\cite{Forn-Diaz10,Wang20}. Recently, another type of light-matter interaction, namely dissipative coupling, was discovered~\cite{Peng16,Metelmann15,Xu16,Wang20-1}. In contrast to the level repulsion of the eigen modes in a coherently coupled system, a dissipatively coupled system is featured by the level attraction. As a new coupling mechanism, the discovery of dissipative coupling develops and broadens various methods for controlling and utilizing light-matter interactions. With the interference between coherent coupling and dissipative coupling, it becomes feasible to engineer nonreciprocal magnonic devices in the classical and quantum regimes~\cite{Wang19,Wang22}. In addition, dissipative coupling also has important applications in, e.g., nonreciprocal photon transmission and amplification~\cite{Metelmann15}, topological energy transfer~\cite{Xu16}, sensitive detection~\cite{Nair21}, and lowering the threshold power of nonlinear effects~\cite{Nair21-2}.

Coherent coupling makes it possible to generate entanglement between two or more quantum systems~\cite{Horodecki09}. Entanglement lies at the core of quantum mechanics and plays a significant role in quantum information processing~\cite{Ekert91,Grover97,Divincenzo95,Lim06,Roghani18} and quantum communication~\cite{Bennett92,Braunstein98,Wang05-Deng,Gisin07}. The typical entangled states include the Bell state, the Greenberger-Horne-Zeilinger (GHZ) state and the $W$ state~\cite{Horodecki09}. Compared with the GHZ state, the $W$ state is more robust with respect to particle losses~\cite{Dur00}. Due to their intrinsic interest and practical importance, many efforts were devoted to generating the Bell state, the GHZ state and the $W$ state in coherently coupled systems (see, e.g., Refs.~\cite{Brunner14,Aolita15,Zeilinger97,Neumann08,Yang12,Zhang20-Liu,Fang19,Stojanovic20,Yang13-Su,Gangat13}). In previous studies, the entanglement of mixed states was investigated in dissipatively coupled systems~\cite{Gonzalez-Tudela11,Liao15,Facchi16,Yang-Wang21}. Very recently, some schemes were proposed to generate the Bell state via dissipative coupling, with the help of continuous measurement or postselection~\cite{Zhang19-Baranger,Zhan22,Zou21}. However, how to prepare the GHZ state and the $W$ state in a dissipatively coupled system has not been studied yet.

In the present paper, we propose a scheme of generating the tripartite $W$ state via reservoir engineering in the absence of any coherent coupling. The physical system considered here consists of one qubit and two resonators, which are dissipatively coupled through a common reservoir. The dissipative coupling can be described using a Lindblad superoperator with a cooperative jump operator, which is in the form of the linear superposition of the qubit operator and the resonator operators. By carefully designing the jump operator, we find that the $W$ state is a dark state of the dissipatively coupled system. When pumping the qubit with an appropriate drive pulse, the system will evolve from its ground state to the tripartite $W$ state (i.e., a dark state of the system). Note that the above results are also valid if the two resonators are replaced by two qubits, because only both the ground states and the first excited states of the two resonators are involved in preparing the $W$ state (cf. Secs.~\ref{model-B} and \ref{Generating-W-states}).

In our scheme, there is no need to perform measurements and adjust the system parameters. Moreover, the prepared $W$ state decouples from the common reservoir and therefore has a very long lifetime~\cite{Dong12,Yang-Wang21,Zanner22}, i.e., the generated $W$ state is steady rather than transient. With the assistance of local dissipations of subsystems, the pure entangled states, such as the GHZ state, can be also prepared (see, e.g., Refs.~\cite{Reiter16,Sun15}). Different from these works, the proposed scheme is based on the nonlocal dissipation (i.e., the cooperative dissipation) due to the common reservoir, which induces the dissipative couplings among the qubit and the two resenators. Our work provides a scheme for generating tripartite $W$ states via reservoir engineering in dissipatively coupled systems, which is applicable to a wide class of systems including waveguide QED systems~\cite{Gu17,Zhang19-Baranger,Zhan22}, dissipatively coupled spins mediated by a magnetic environment~\cite{Zou21}, magnon-based hybrid systems~\cite{Wang20-1,Wang22}, and so on.

Using this scheme, we further study how to generate the $W$ state in a magnon-based hybrid system. Magnons are collective spin excitations in ferromagnetic crystals~\cite{Gurevich96,Xiong22}. The magnon-based hybrid systems have recently become a promising platform for quantum technologies~\cite{Quirion19,Rameshti21,Yuan22}. Recently, there were some investigations on quantum entanglement in magnon-based hybrid systems~\cite{Li18,Yu20,Yuan20,Sun21}. Specifically, a protocol for producing transient Bell states and GHZ states in a hybrid qubit-photon-magnon system was presented in Ref.~\cite{Qi22}, but how to generate the $W$ state in a magnon-based hybrid system is still an open question. On the other hand, the exotic effects induced by dissipative coupling were widely studied in magnon-based hybrid systems, both theoretically~\cite{Wang22,Nair21,Nair21-2,Grigoryan18,Yu19} and experimentally~\cite{Wang19,Harder18,Li22,Bhoi19}. Motivated by these works, we will apply our results to generate a hybrid $W$ state in a qubit-photon-magnon system, where a superconducting transmon qubit, a superconducting transmission-line resonator and a yttrium iron garnet (YIG) sphere are dissipatively coupled through a coplanar waveguide. The presence of the intrinsic dissipations of the qubit, the resonator and the magnon mode in the YIG sphere limits the lifetime of the hybrid $W$ state. This underpins the utility of our scheme for generating tripartite $W$ states in dissipatively coupled systems. In the versatile magnon-based quantum information processing platform~\cite{Quirion19,Rameshti21,Yuan22}, superconducting qubits (superconducting resonators) can act as quantum processor (quantum bus)~\cite{Xiang13}, while magnon modes can play the role of quantum memory~\cite{Tanji09,Zhang15-Zou}. Using the qubit-photon-magnon $W$ state, quantum processor, quantum bus and quantum memory can be connected. In addition, we can also use the hybrid $W$ state to transfer quantum states among quantum processor, quantum bus and quantum memory~\cite{Bin22}.

Our paper is structured as follows. In Sec.~\ref{model}, we describe the dissipatively coupled ternary system and give the corresponding Lindblad master equation. Using the master equation, we investigate the dark states and the bright states of the dissipatively coupled system. In Sec.~\ref{Generating-W-states}, we demonstrate that the $W$ state is exactly a dark state of the system. We further show how to generate the $W$ state by pumping the qubit with an appropriate drive pulse. In Sec.~\ref{qubit-photon-magnon}, we apply our results to create the $W$ state in a hybrid qubit-photon-magnon system. A brief summary is given in Sec.~\ref{conclusions}.

\begin{figure}
\includegraphics[width=0.45\textwidth]{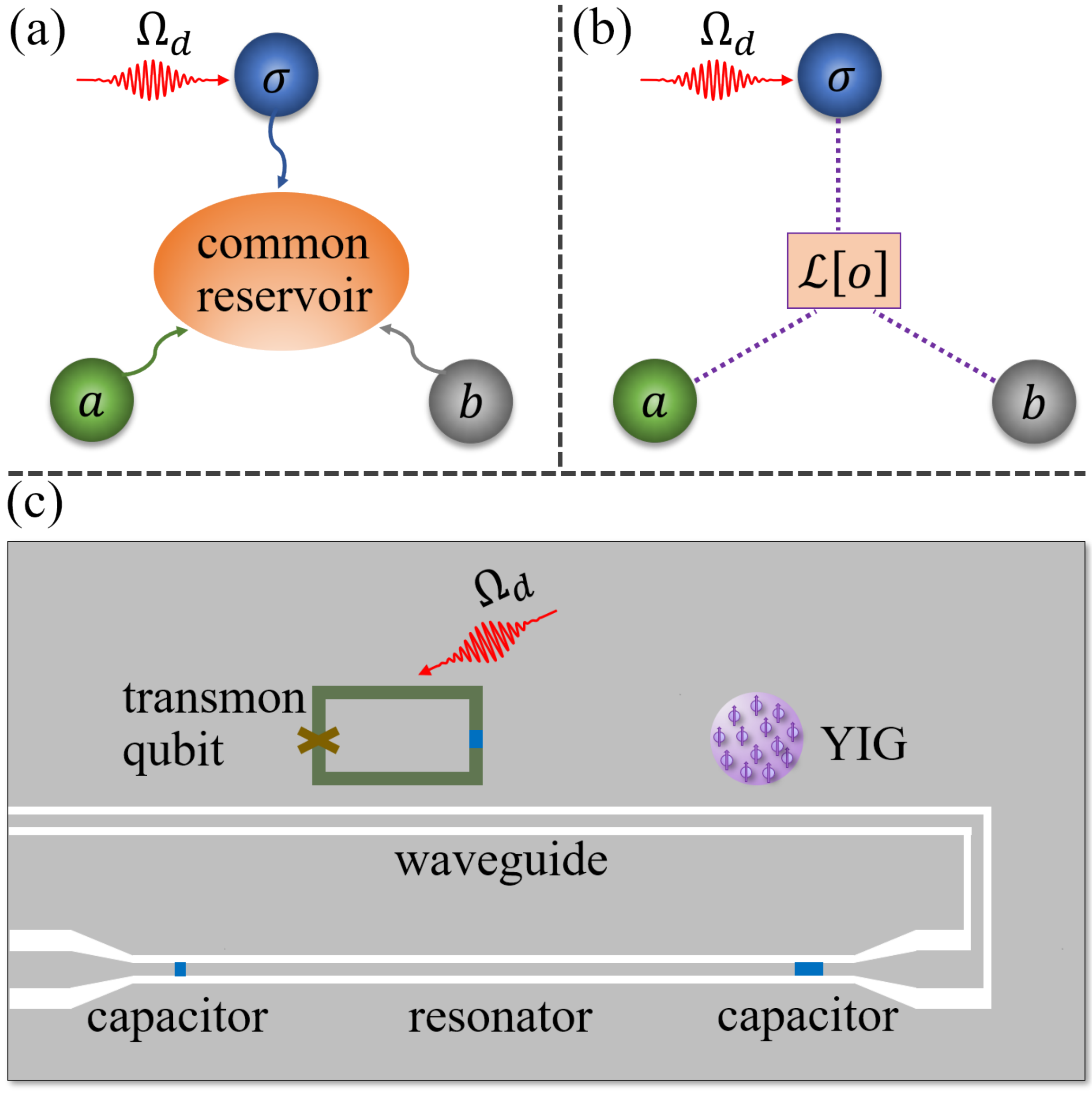}
\caption{(a) A schematic showing a qubit and two resonators (with annihilation operators $\sigma$, $a$ and $b$, respectively) simultaneously interacting with a common reservoir. To generate the $W$ state, a drive pulse with Rabi frequency $\Omega_d$ pumps the qubit. (b) Mediated by the common reservoir in (a), the three subsystems are dissipatively coupled together, which can be described using a Lindblad superoperator $\mathcal{L}[o]$ in the master equation, cf. Eqs.~(\ref{master}) to (\ref{jump-operator}). (c) Diagram of a hybrid qubit-photon-magnon system. A superconducting transmon qubit, a superconducting transmission-line resonator and a ferrimagnetic YIG sphere are dissipatively coupled via an open waveguide, where the transmon qubit is driven by a microwave pulse with Rabi frequency $\Omega_d$.}
\label{fig1}
\end{figure}

\section{Model}\label{model}

\subsection{Master equation}

As depicted in Fig.~\ref{fig1}(a), we consider a general model for a qubit and two resonators, which simultaneously couple to a common reservoir of many bosonic modes. In the absence of the drive field on the qubit, the Hamiltonian of the ternary system can be written as (setting $\hbar=1$):
\begin{eqnarray}\label{Hamiltonian}
H_s&=&\omega_\sigma \sigma^\dag \sigma + \omega_a a^\dag a + \omega_b b^\dag b,
\end{eqnarray}
where $\omega_\sigma$ is the frequency of the qubit, $\sigma = |g\rangle {\langle e|}$ ($\sigma^\dag=|e\rangle \langle g|$) is the lowering (raising) operator of the qubit with ground state $|g\rangle$ and excited state $|e\rangle$, $a$ and $a^\dag$ ($b$ and $b^\dag$) are the annihilation and creation operators of resonator $a$ (resonator $b$) at frequency $\omega_a$ ($\omega_b$). When the three subsystems are nearly resonant, i.e., $\omega_\sigma \approx \omega_a \approx \omega_b  \approx \omega_0$, by tracing out over the degrees of freedom of the reservoir, the master equation for the density operator $\rho$ of the ternary system can be derived as (see the Appendix~\ref{Appendix-A}):
\begin{equation}\label{master}
\dot{\rho}=-i[H_s,\rho]+\tau \mathcal{L}[o]\rho,
\end{equation}
where $\tau=\sum_k \pi \lambda_k^2\delta(\omega_0-\omega_k)$ is the decay rate of the ternary system, and $\lambda_{k}$ is the effective coupling strength between the ternary system and the $k$th mode of the reservoir with frequency $\omega_k$. Note that in the derivations for the above master equation, the Born-Markovian approximation and the assumption of the reservoir at zero temperature were used. The Lindblad superoperator $\mathcal{L}[o]\rho$ describes the dissipative interactions among the three subsystems mediated by the common reservoir [cf. Fig.~\ref{fig1}(b)], which is given by
\begin{equation}\label{superoperator}
\mathcal{L}[o]\rho=2o\rho o^\dag-o^\dag o\rho-\rho o^\dag o,
\end{equation}
with the jump operator
\begin{eqnarray}\label{jump-operator}
o=\eta_\sigma \sigma+\eta_a a+\eta_b b,
\end{eqnarray}
where $\eta_\alpha=(\lambda_{k\alpha}/\lambda_{k})e^{-i\phi_\alpha}$ ($\alpha=\sigma,a,b$), $\lambda_{k\alpha}$ is the coupling strength between the subsystem $\alpha$ and the $k$th mode of the reservoir, and the corresponding phase is $\phi_\alpha=(\omega_0/\upsilon)x_\alpha$ with the speed $\upsilon$ of light and the location $x_\alpha$ of the subsystem $\alpha$. It should be emphasized that the cooperative dissipative term $\tau \mathcal{L}[o]\rho$ in Eq.~(\ref{master}) results from the interaction of the entire ternary system with a common reservoir. If two subsystems (e.g., the two resonators) of the ternary system decouple from the common reservoir (i.e., $\eta_a = \eta_b = 0$ but $\eta_\sigma \neq 0$), the dissipative term $\tau \mathcal{L}[o]\rho$ will be reduced to $(|\eta_\sigma|^2\tau) \mathcal{L}[\sigma]\rho$, which presents the local dissipation of the qubit.

For clarity, we can expand the Lindblad superoperator in Eq.~(\ref{superoperator}) and obtain
$\mathcal{L}[o]\rho=\sum_{\alpha\alpha'}\eta_\alpha\eta_{\alpha'}^*(2\alpha\rho \alpha'^\dag-\alpha'^\dag \alpha\rho-\rho \alpha'^\dag \alpha)$. The diagonal term (with $\alpha=\alpha'$) presents the local dissipation of the subsystem $\alpha$ with the decay rate $|\eta_\alpha|^2\tau$. In general, $\eta_\alpha\eta_{\alpha'}^*=(\lambda_{k\alpha}\lambda_{k\alpha'}/\lambda_k^2)e^{-i(\phi_\alpha-\phi_{\alpha'})}$
is a complex number due to the phase $\phi_\alpha-\phi_{\alpha'}$. Note that we assumed $\lambda_{k\alpha}$ and $\lambda_k$ are real. Therefore, the off-diagonal term (with $\alpha \neq \alpha'$) denotes both the dissipative coupling with strength ${\rm Re}[\eta_\alpha\eta_{\alpha'}^*]\tau$~\cite{Metelmann15,Wang20-1} and the coherent coupling with strength ${\rm Im}[\eta_\alpha\eta_{\alpha'}^*]\tau$~\cite{Gonzalez-Tudela11,Martin-Cano11} between the two subsystems $\alpha$ and $\alpha'$. In our paper, we only study the case of $\phi_\alpha-\phi_{\alpha'}=\pm n\pi$ ($n=0,1,2,...,$) by setting $x_\alpha-x_{\alpha'}=\pm n\pi\upsilon/\omega_0$, i.e., there is not any coherent coupling among the three subsystems because of ${\rm Im}[\eta_\alpha\eta_{\alpha'}^*]\tau=0$.

\subsection{Dark states of the system}\label{model-B}

For the dissipatively coupled system, there are some dark states, which are decoupled from the common reservoir and have long lifetimes~\cite{Dong12,Zanner22}. To study the dark states of the system, we rewrite the master equation (\ref{master}) in the form~\cite{Minganti19,Zhang21}: $\dot{\rho}=-i(H_{\rm eff}\rho-\rho H^\dag_{\rm eff})+2\tau o\rho o^\dag$, where
\begin{equation}\label{non-Hermitian}
H_{\rm eff}=\omega_\sigma \sigma^\dag \sigma + \omega_a a^\dag a + \omega_b b^\dag b-i\tau o^\dag o
\end{equation}
is an effective non-Hermitian Hamiltonian. Obviously, the Hamiltonian $H_{\rm eff}$ preserves the total number $\mathcal{N}$ of excitations in the system due to $[H_{\rm eff},\mathcal{N}]=0$, with
\begin{equation}\label{number}
\mathcal{N}= \sigma^\dag \sigma + a^\dag a +  b^\dag b.
\end{equation}
Thus, we can analyze the eigenvectors of the dissipatively coupled system in the closed one-excitation subspace $\{|e00\rangle,\,\,|g10\rangle,\,\,|g01\rangle\}$ with $|e00\rangle \equiv |e\rangle|0\rangle_a|0\rangle_b$, $|g10\rangle \equiv |g\rangle|1\rangle_a|0\rangle_b$ and $|g01\rangle\equiv |g\rangle|0\rangle_a|1\rangle_b$, where $|n\rangle_{a\,(b)}$ is the Fock state of the resonator $a\,\,(b)$, and $n$ is the corresponding excitation number in the resonator $a$ ($b$). When the three subsystems are resonant, i.e., $\omega_\sigma = \omega_a = \omega_b  = \omega_0$, the non-Hermitian Hamiltonian $H_{\rm eff}$ in Eq.~(\ref{non-Hermitian}) has three orthonormal eigenvectors in the one-excitation subspace, two degenerate dark states $\{|D_1\rangle,\,\,|D_2\rangle\}$ and one bright state $|B\rangle$, which are given by
\begin{eqnarray}\label{eigenvectors}
|D_1\rangle&=&\frac{1}{\sqrt{|\eta_\sigma|^2+|\eta_b|^2}}\left(\eta_b|e00\rangle-\eta_\sigma|g01\rangle\right),\nonumber\\
|D_2\rangle&=&\frac{1}{\sqrt{|k_1|^2+|k_2|^2+|k_3|^2}}\left(k_1|e00\rangle+k_2|g10\rangle+k_3|g01\rangle\right),\nonumber\\
|B\rangle&=&\frac{1}{\sqrt{|\eta_\sigma|^2+|\eta_a|^2+|\eta_b|^2}}
                  \left(\eta_\sigma^*|e00\rangle+\eta_a^*|g10\rangle+\eta_b^*|g01\rangle\right),\nonumber\\
\end{eqnarray}
with
\begin{equation}\label{}
k_1=\frac{|\eta_\sigma|^2\eta_a}{|\eta_\sigma|^2+|\eta_b|^2},~~~k_2=-\eta_\sigma,
~~~k_3=\frac{\eta_\sigma\eta_a\eta_b^*}{|\eta_\sigma|^2+|\eta_b|^2}.
\end{equation}
Note that in Eq.~(\ref{eigenvectors}), the Gram-Schmidt orthogonalization was applied. The corresponding eigenvalues for $\{|D_1\rangle,\,|D_2\rangle,\,|B\rangle\}$ are $E_{\rm D1}=E_{\rm D2}=\omega_0$ and $E_{B}=\omega_0-i(|\eta_\sigma|^2+|\eta_a|^2+|\eta_b|^2)\tau$, respectively. From these eigenvectors and eigenvalues, we can know that the two dark states $\{|D_1\rangle,\,\,|D_2\rangle\}$ are stable because their decay rates are zero, while the bright state $|B\rangle$ is a super-radiant state with decay rate $(|\eta_\sigma|^2+|\eta_a|^2+|\eta_b|^2)\tau$.

\begin{figure}
\includegraphics[width=0.45\textwidth]{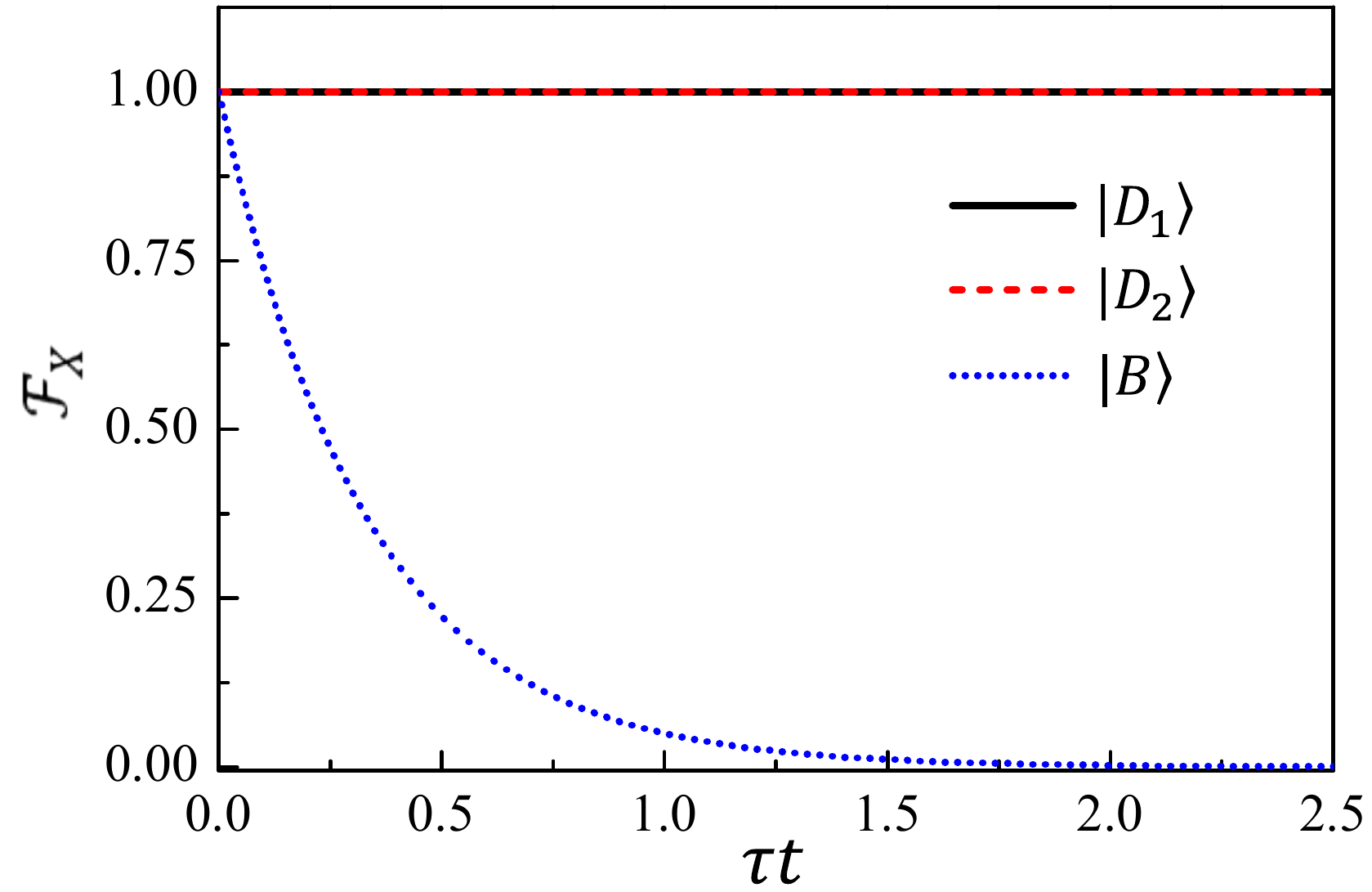}
\caption{Time evolution of the fidelity $\mathcal{F}_X={\rm Tr}(\rho |X\rangle\langle X|)$ for the initial states $|X\rangle=|D_1\rangle$ (black solid curve), $|X\rangle=|D_2\rangle$ (red dashed curve), and $|X\rangle=|B\rangle$ (blue dotted curve), calculated using Eq.~(\ref{master}). Here the parameters are set as $\omega_\sigma/\tau=\omega_a/\tau=\omega_b/\tau=500$ and $\eta_\sigma=\eta_a=\eta_b=1$.}
\label{fig2}
\end{figure}

By numerically solving the master equation in Eq.~(\ref{master}), we study the stability of the eigenvectors $\{|D_1\rangle,\,|D_2\rangle,\,|B\rangle\}$. As shown in Fig.~\ref{fig2}, if the initial state of the system is the dark state $|D_1\rangle$ or $|D_2\rangle$, the fidelity $\mathcal{F}_{D_1}={\rm Tr}(\rho |D_1\rangle\langle D_1|)$ or $\mathcal{F}_{D_2}={\rm Tr}(\rho |D_2\rangle\langle D_2|)$ is independent of time $t$ (i.e., $\mathcal{F}_{D_1}=\mathcal{F}_{D_2}=1$, see the black solid and red dashed curves), which indicates that the two dark states $|D_1\rangle$ and $|D_2\rangle$ are stable. However, when the system is in the bright state $|B\rangle$ at $t=0$, the state of the system will decay with time $t$ to the vacuum state $|g00\rangle$, and the corresponding fidelity $\mathcal{F}_B={\rm Tr}(\rho |B\rangle\langle B|)$ decreases monotonically from 1 to 0 (see the blue dotted curve).

We would like to point out that the assumption of a one-excitation subspace in Eq.~(\ref{eigenvectors}) does not limit the generality of our work. In Secs.~\ref{Generating-W-states} and \ref{qubit-photon-magnon}, the numerical results related to generating $W$ states are obtained via solving the master equation of the system in a large-enough Hilbert space (rather than the subspace $\{|g00\rangle,|e00\rangle,|g10\rangle,|g01\rangle\}$), which are almost consistent with the theoretical analyses in the one-excitation subspace. The reason is that since the drive on the system is weak and the bright state $|B\rangle$ decays fast, the higher-excitation subspace is nearly not occupied in preparing $W$ states. Therefore, the assumption of the one-excitation subspace applied in our theoretical model is reasonable.

\section{Generating long-lived $W$ states}\label{Generating-W-states}

In the dissipatively coupled system, any single-excitation state can be expressed as a linear superposition of the three orthonormal eigenvectors $\{|D_1\rangle,\,|D_2\rangle,\,|B\rangle\}$ given in Eq.~(\ref{eigenvectors}). For example, the state $|e00\rangle$ can be expressed as
\begin{equation}\label{e00}
|e00\rangle=c_{1}|D_1\rangle+c_2|D_2\rangle+c_3|B\rangle,
\end{equation}
where the coefficients are $c_1=\langle D_1 |e00\rangle$, $c_2=\langle D_2 |e00\rangle$ and  $c_3=\langle B |e00\rangle$, i.e.,
\begin{eqnarray}\label{}
c_1&=&\frac{\eta_b^*}{\sqrt{|\eta_\sigma|^2+|\eta_b|^2}},\nonumber\\
c_2&=&\frac{k_1^*}{\sqrt{|k_1|^2+|k_2|^2+|k_3|^2}},\nonumber\\
c_3&=&\frac{\eta_\sigma}{\sqrt{|\eta_\sigma|^2+|\eta_a|^2+|\eta_b|^2}}.
\end{eqnarray}
If the ternary system is prepared in the state $|e00\rangle$ at $t=0$, the components of the dark states $\{|D_1\rangle,\,|D_2\rangle\}$ are stable, while the component of the bright mode $|B\rangle$ will decay with time $t$ to the vacuum state $|g00\rangle$. Finally, the system reaches a steady mixed state of the vacuum state $|g00\rangle$ and the dark state
\begin{eqnarray}\label{dark}
|D\rangle&=&\frac{1}{\sqrt{|c_1|^2+|c_2|^2}}\left(c_1|D_1\rangle+c_2|D_2\rangle\right).
\end{eqnarray}
With appropriate values of the parameters $(\eta_\sigma,\, \eta_a,\,\eta_b)$, the dark state $|D\rangle$ can be an arbitrary $W$ state with single excitation (cf. Table~\ref{Table}). For $\eta_\sigma=2$ and $\eta_a=\eta_b=-1$, the corresponding dark state $|D\rangle$ is the prototype $W$ state $|W_3^{(1)}\rangle=(|e00\rangle+|g10\rangle+|g01\rangle)/\sqrt{3}$. In addition, we can also obtain the Agrawal $W$ state $|W_3^{(2)}\rangle=(\sqrt{2}|e00\rangle+|g10\rangle+|g01\rangle)/2$ in the case of $\eta_\sigma=\sqrt{2}$ and $\eta_a=\eta_b=-1$. The Agrawal $W$ state $|W_3^{(2)}\rangle$ was first proposed by Agrawal and Pati, which has important applications in perfect teleportation and superdense coding~\cite{Agrawal06}.

\newcommand{\tabincell}[2]{}
\renewcommand\tabcolsep{25.5pt}
\begin{table}
	\centering
	\scriptsize
	\caption{The form of the dark state $|D\rangle$ given in Eq.~(\ref{dark}) for different values of the parameters $(\eta_\sigma,\,\eta_a,\,\eta_b)$.}
	\label{tab:notations}
	\begin{tabular}{cc}
		\\[-1mm]
		\hline
		\hline\\[-1mm]
		 $(\eta_\sigma,\,\eta_a,\,\eta_b)$ &\qquad Dark state $|D\rangle$\\[1mm]
		\hline
		\vspace{-3mm}\\[2mm]
		$(2,\,-1,\,-1)$  & $|W_3^{(1)}\rangle=(|e00\rangle+|g10\rangle+|g01\rangle)/\sqrt{3}$\\[2mm]
		$(\sqrt{2},\,-1,\,-1)$  & $|W_3^{(2)}\rangle=(\sqrt{2}|e00\rangle+|g10\rangle+|g01\rangle)/2$\\[2mm]
        $(1,\,1,\,1)$  & $|W_3^{(3)}\rangle=(2|e00\rangle-|g10\rangle-|g01\rangle)/\sqrt{6}$\\[2mm]
        $(1,\,0,\,-1)$  & $|W_3^{(4)}\rangle=(|e00\rangle+|g01\rangle)/\sqrt{2}$\\[2mm]
        \hline
		\hline
	\end{tabular}
\label{Table}
\end{table}

\begin{figure}
\includegraphics[width=0.48\textwidth]{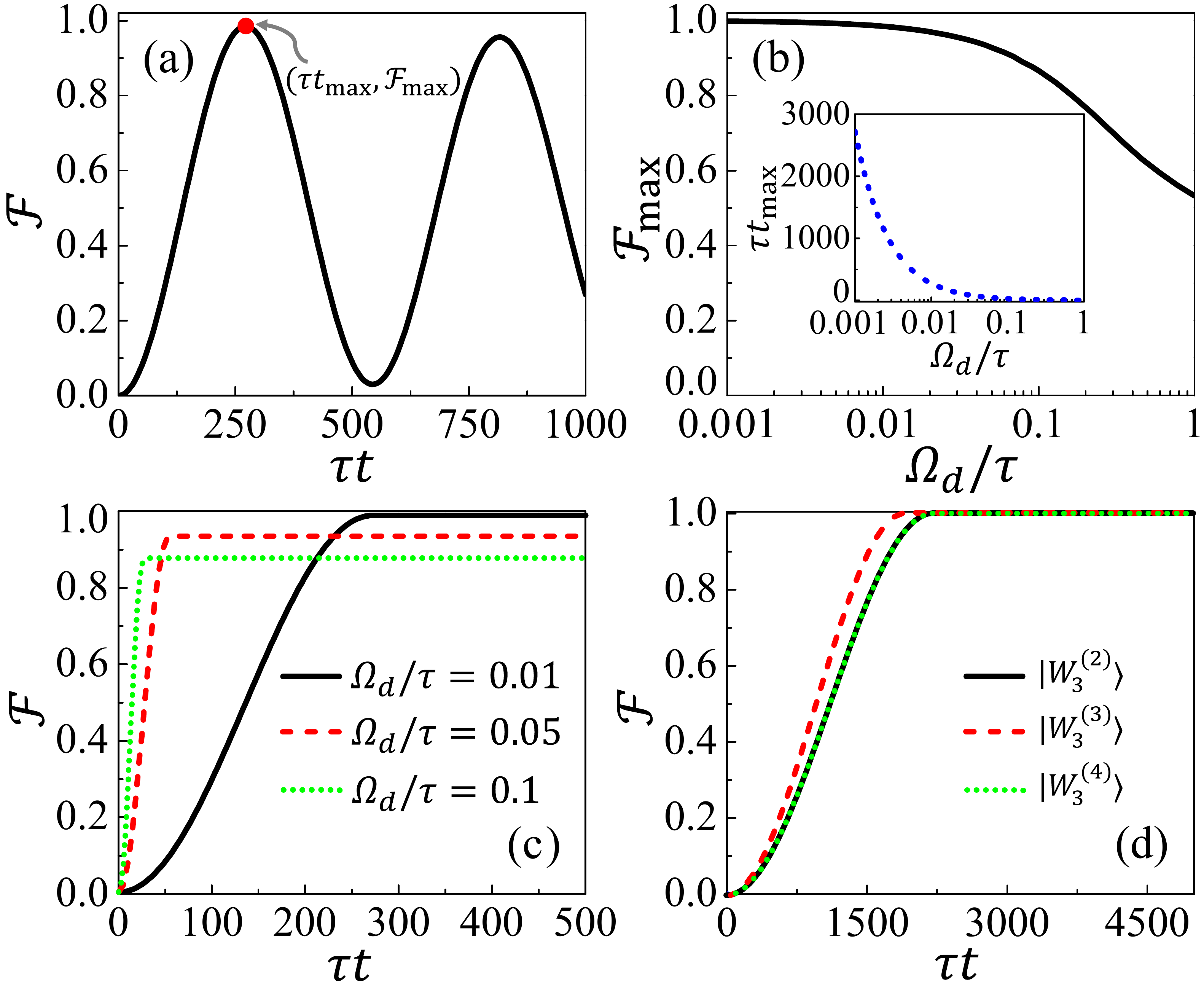}
\caption{(a) Time evolution of the fidelity $\mathcal{F}$ of the prototype $W$ state $|W_3^{(1)}\rangle=(|e00\rangle+|g10\rangle+|g01\rangle)/\sqrt{3}$, with $\Omega_d/\tau=0.01$ and $t_0 = +\infty$, where the red dot denotes the maximum value $\mathcal{F}_{\rm max}=0.985$ of the fidelity $\mathcal{F}$ at time $\tau t_{\rm max}=273$. (b) The maximal fidelity $\mathcal{F}_{\rm max}$ of $|W_3^{(1)}\rangle$ versus the Rabi frequency $\Omega_d/\tau$, where the corresponding time $\tau t_{\rm max}$ (for reaching $\mathcal{F}_{\rm max}$) versus $\Omega_d/\tau$ is shown in the inset. (c) Time evolution of the fidelity $\mathcal{F}$ of $|W_3^{(1)}\rangle$ for drive pulses with different shapes: $\Omega_d/\tau=0.01$, $\tau t_0=273$ (black solid curve); $\Omega_d/\tau=0.05$, $\tau t_0=54.1$ (red dashed curve); and  $\Omega_d/\tau=0.1$, $\tau t_0=26.9$ (green dotted curve). (d) Time evolution of the fidelity $\mathcal{F}$ for different $W$ states: the Agrawal $W$ state $|W_3^{(2)}\rangle=(\sqrt{2}|e00\rangle+|g10\rangle+|g01\rangle)/2$ with $\Omega_d/\tau=0.001$, $\tau t_0=2211.1$ (black solid curve); the common $W$ state $|W_3^{(3)}\rangle=(2|e00\rangle-|g10\rangle-|g01\rangle)/\sqrt{6}$ with $\Omega_d/\tau=0.001$, $\tau t_0=1934.7$ (red dashed curve); and the Bell state $|W_3^{(4)}\rangle=(|e00\rangle+|g01\rangle)/\sqrt{2}$ with $\Omega_d/\tau=0.001$, $\tau t_0/=2211.1$ (green dotted curve). These results in (a)-(d) are obtained using the master equation in Eq.~(\ref{master}) with the system Hamiltonian $H_s^{(d)}$ given in Eq.~(\ref{Hamiltonian-drive}) and the initial state $|g00\rangle$. For the four $W$ states, $|W_3^{(1)}\rangle$, $|W_3^{(2)}\rangle$, $|W_3^{(3)}\rangle$ and $|W_3^{(4)}\rangle$, the corresponding values of $(\eta_\sigma,\,\eta_a,\,\eta_b)$ can be found in Table~\ref{Table}. Other parameters are $\omega_\sigma/\tau=\omega_a/\tau=\omega_b/\tau=\omega_d/\tau=500$.}
\label{fig3}
\end{figure}

At initial time $t=0$, the system is in the ground state $|g00\rangle$. To produce the single-excitation $W$ state (i.e., the dark state $|D\rangle$), we can use a classical coherent drive pulse with frequency $\omega_d$ and duration $t_0$ to pump the qubit. This corresponds to adding the drive term $\Theta(t_0-t)\Omega_d(\sigma^\dag e^{-i\omega_d t}+\sigma e^{i\omega_d t})$ to the system Hamiltonian $H_s$ in Eq.~(\ref{Hamiltonian}), which is then expressed as
\begin{eqnarray}\label{Hamiltonian-drive}
H_s^{(d)}&=&\omega_\sigma \sigma^\dag \sigma + \omega_a a^\dag a + \omega_b b^\dag b\nonumber\\
   & &+\Theta(t_0-t)\Omega_d(\sigma^\dag e^{-i\omega_d t}+\sigma e^{i\omega_d t}),
\end{eqnarray}
where $\Omega_d$ is the Rabi frequency, and $\Theta(t_0-t)$ is the Heaviside function. Now the dynamics of the ternary system is also governed by the master equation in Eq.~(\ref{master}), but $H_s$ is replaced by $H_s^{(d)}$, where we neglect the effect of the drive field on the dissipative term $\tau \mathcal{L}[o]\rho$ in Eq.~(\ref{master}). This approximation is reasonable for a weak drive field, which is widely used in quantum optics~\cite{Scully97}. Correspondingly, the effective non-Hermitian Hamiltonian $H_{\rm eff}$ in Eq.~(\ref{non-Hermitian}) becomes $H_{\rm eff}^{(d)}=H_{\rm eff}+\Theta(t_0-t)\Omega_d(\sigma^\dag e^{-i\omega_d t}+\sigma e^{i\omega_d t})$. Obviously, the presence of the drive term spoils the preservation of the total number of excitations of the system due to $[H_{\rm eff}^{(d)},\mathcal{N}]\neq 0$ when $t<t_0$, where the total number $\mathcal{N}$ of excitations in the ternary system is given in Eq.~(\ref{number}). As a result, the one-excitation subspace of the system is not closed. Thus, it is difficult to exactly obtain the eigenvectors of $H_{\rm eff}^{(d)}$. Fortunately, since the drive field is very weak (i.e., $\Omega_d \ll \tau$) in our scheme, we can approximatively assume that the three eigenvectors of $H_{\rm eff}$ in Eq.~(\ref{eigenvectors}) are also the eigenvectors of $H_{\rm eff}^{(d)}$ in the one-excitation subspace. For $t>t_0$, the drive pulse ends, and $H_{\rm eff}^{(d)}=H_{\rm eff}$.

With the system Hamiltonian $H_s^{(d)}$ in Eq.~(\ref{Hamiltonian-drive}), we plot the time evolution of the fidelity $\mathcal{F}$ of the prototype $W$ state $|W_3^{(1)}\rangle$ in Fig.~\ref{fig3}(a) via numerically solving the master equation in Eq.~(\ref{master}) with $H_s$ replaced by $H_s^{(d)}$, where the fidelity $\mathcal{F}$ of the target $W$ state (i.e., the dark state $|D\rangle$) is defined as
\begin{equation}\label{}
\mathcal{F}={\rm Tr}\left(\rho|D\rangle\langle D|\right).
\end{equation}
Here we assume that the pulse length is infinite, i.e., $t_0 = +\infty$. When $\tau t< \tau t_{\rm max}$ (with $\tau t_{\rm max}=273)$, the fidelity $\mathcal{F}$ increases monotonically from $\mathcal{F}=0$ at $t=0$ to its maximum value $\mathcal{F}=\mathcal{F}_{\rm max}$ (with $\mathcal{F}_{\rm max}=0.985$) at time $\tau t= \tau t_{\rm max}$. The corresponding physical mechanisms are as follows. (i) When the qubit is pumped by the drive pulse, the qubit is excited and the state $|g00\rangle$ of the system transfers to $|e00\rangle$. (ii) Due to the dissipative couplings, the state $|e00\rangle$ will evolve into the mixed state of the dark state $|D\rangle$ (i.e., the prototype $W$ state $|W_3^{(1)}\rangle$) and the ground state $|g00\rangle$. (iii) For the component $|g00\rangle$ in the mixed state, it will be pumped into the state $|e00\rangle$ again. (iv) The processes (ii) and (iii) are repeated. Strictly speaking, now the prototype $W$ state $|W_3^{(1)}\rangle$ is not the dark state of the system, because the effective non-Hermitian Hamiltonian in Eq.~(\ref{non-Hermitian}) does not include the drive term. This results in that the maximum value $\mathcal{F}_{\rm max}$ of the fidelity $\mathcal{F}$ is smaller than the ideal value 1 (i.e., $\mathcal{F}_{\rm max}<1$) and that the fidelity $\mathcal{F}$ versus time $t$ exhibits obvious oscillations [cf. Fig.~\ref{fig3}(a)]. Using a weaker drive pulse, we can obtain a higher fidelity $\mathcal{F}_{\rm max}$ at the expense of a larger $t_{\rm max}$ for reaching $\mathcal{F}_{\rm max}$ [cf. Fig.~\ref{fig3}(b)]. In the weak-drive limit $\Omega_d \rightarrow 0$, $\mathcal{F}_{\rm max} \rightarrow 1$ but $t_{\rm max} \rightarrow +\infty$ (e.g., when $\Omega_d/\tau=0.001$, $\mathcal{F}_{\rm max} = 0.998$ and $\tau t_{\rm max} =2720.7$).

For generating a stable $W$ state with fidelity $\mathcal{F}=\mathcal{F}_{\rm max}$, the duration $t_0$ of the pulse should be equal to $t_{\rm max}$ (i.e., $t_0=t_{\rm max}$). In Fig.~\ref{fig3}(c), we present the time evolution of the fidelity $\mathcal{F}$ of the prototype $W$ state $|W_3^{(1)}\rangle$ for different values of Rabi frequency $\Omega_d$, where the corresponding duration $t_0=t_{\rm max}$ of the drive pulse can be found in Fig.~\ref{fig3}(b). As we expected, the fidelity $\mathcal{F}$ becomes time independent after reaching the maximum value $\mathcal{F}_{\rm max}$ at time $t=t_0$, i.e., the lifetime of the prototype $W$ state $|W_3^{(1)}\rangle$ is infinite. When the drive pulse becomes weaker, the fidelity of the prepared $|W_3^{(1)}\rangle$ is higher, while the time for preparing stable $|W_3^{(1)}\rangle$ is longer. Moreover, as shown in Fig.~\ref{fig3}(d), we can also generate other types of stable $W$ states by selecting specific values of $(\eta_\sigma,\,\eta_a,\,\eta_b)$, such as the Agrawal $W$ state $|W_3^{(2)}\rangle=(\sqrt{2}|e00\rangle+|g10\rangle+|g01\rangle)/2$ (black solid curve), the common $W$ state $|W_3^{(3)}\rangle=(2|e00\rangle-|g10\rangle-|g01\rangle)/\sqrt{6}$ (red dashed curve), and the Bell state $|W_3^{(4)}\rangle=(|e00\rangle+|g01\rangle)/\sqrt{2}$ (green dotted curve).

\section{Generating $W$ states via dissipative couplings in a hybrid qubit-photon-magnon system}\label{qubit-photon-magnon}

Our results provide a way to produce $W$ states in dissipatively coupled systems. Guided by the recent experiments related to dissipative coupling in magnon-based hybrid systems~\cite{Wang19,Harder18,Li22,Bhoi19}, we apply the proposed scheme in a hybrid qubit-photon-magnon system. As depicted in Fig.~\ref{fig1}(c), the hybrid system comprises a superconducting transmon qubit, a superconducting transmission-line resonator and a YIG sphere, which are simultaneously coupled to an open waveguide. Owing to the common reservoir (i.e., the waveguide), there are dissipative couplings among the three quantum subsystems. In addition, we use a microwave pulse to drive the transmon qubit for generating the $W$ states. Now the total Hamiltonian of the qubit-photon-magnon system can be cast exactly in the form of Eq.~(\ref{Hamiltonian-drive}) and the dynamics of the hybrid system is governed by the Lindblad master equation in Eq.~(\ref{master}), where the microwave photon operator is $a$ and the magnon operator is $b$.

\begin{figure}
\includegraphics[width=0.48\textwidth]{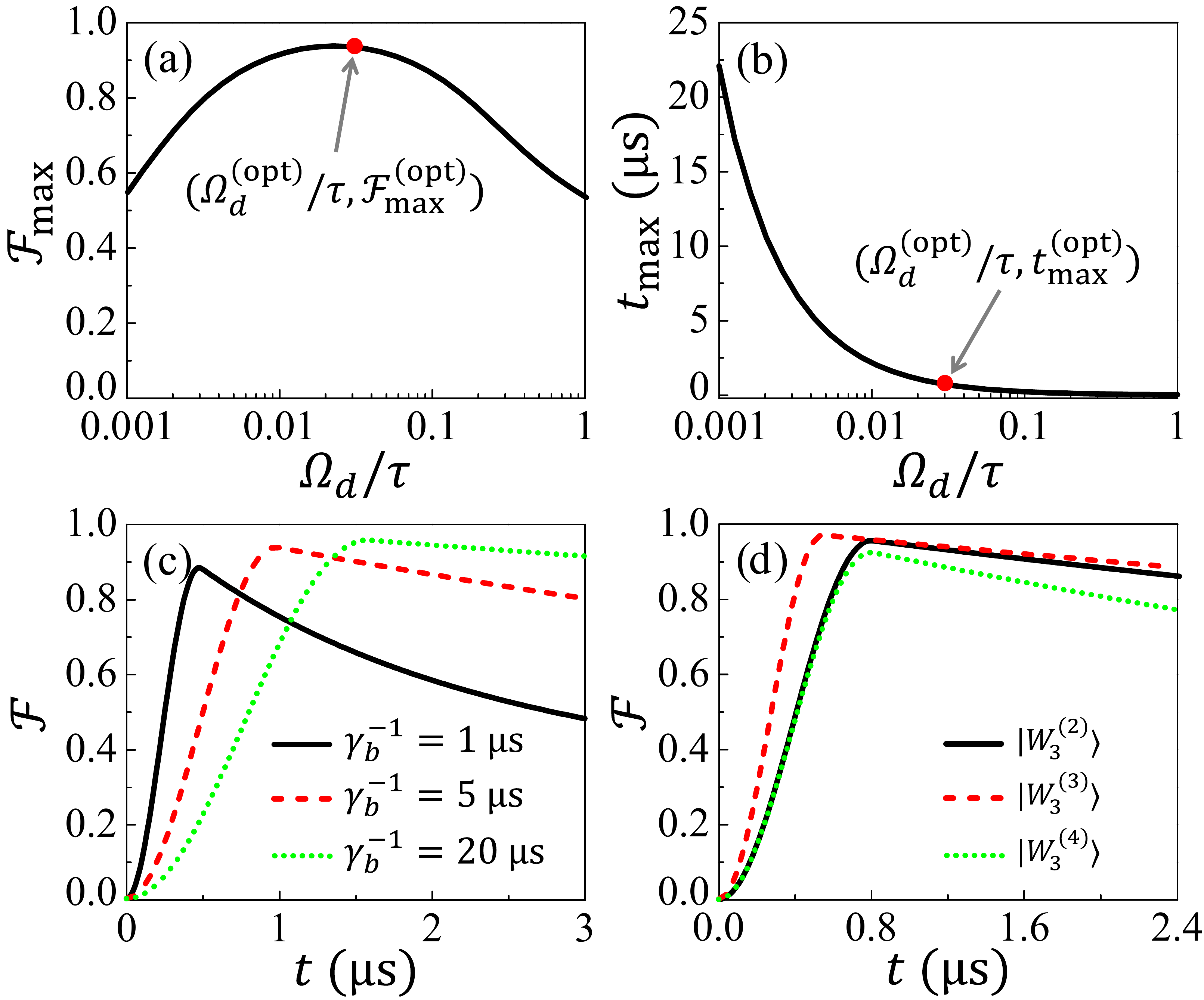}
\caption{(a) The maximal fidelity $\mathcal{F}_{\rm max}$ of the prototype $W$ state $|W_3^{(1)}\rangle=(|e00\rangle+|g10\rangle+|g01\rangle)/\sqrt{3}$ versus the Rabi frequency $\Omega_d/\tau$ with $\gamma_b^{-1}=5\,\,{\rm\mu s}$, where the corresponding time $t_{\rm max}$ (for reaching $\mathcal{F}_{\rm max}$) versus $\Omega_d/\tau$ is shown in (b). (c) Time evolution of the fidelity $\mathcal{F}$ of $|W_3^{(1)}\rangle$ for different relaxation times of the magnon mode: $\gamma_b^{-1}=1\,\,{\rm\mu s}$ (black solid curve), $\gamma_b^{-1}=5\,\,{\rm\mu s}$ (red dashed curve), and $\gamma_b^{-1}=20\,\,{\rm\mu s}$ (green dotted curve). The shapes of the corresponding drive pulses are $\Omega_d/\tau=0.0452$, $t_0=0.478\,\,{\rm\mu s}$; $\Omega_d/\tau=0.0221$, $t_0=0.976\,\,{\rm\mu s}$; and $\Omega_d/\tau=0.0137$, $t_0=1.572\,\,{\rm\mu s}$, respectively. (d) Time evolution of the fidelity $\mathcal{F}$ for different $W$ states in the case of $\gamma_b^{-1}=5\,\,{\rm\mu s}$: the Agrawal $W$ state $|W_3^{(2)}\rangle=(\sqrt{2}|e00\rangle+|g10\rangle+|g01\rangle)/2$ with $\Omega_d/\tau=0.0221$, $t_0=0.795\,\,{\rm\mu s}$ (black solid curve); the common $W$ state $|W_3^{(3)}\rangle=(2|e00\rangle-|g10\rangle-|g01\rangle)/\sqrt{6}$ with $\Omega_d/\tau=0.0281$, $t_0=0.547\,\,{\rm\mu s}$ (red dashed curve); and the Bell state $|W_3^{(4)}\rangle=(|e00\rangle+|g01\rangle)/\sqrt{2}$ with $\Omega_d/\tau=0.0221$, $t_0=0.795\,\,{\rm\mu s}$ (green dotted curve). These results in (a)-(d) are obtained using the master equation in Eq.~(\ref{master-magnon}) with the system Hamiltonian $H_s^{(d)}$ given in Eq.~(\ref{Hamiltonian-drive}) and the initial state $|g00\rangle$. Other parameters are $\gamma_\sigma^{-1}=\gamma_a^{-1}=60\,\,{\rm\mu s}$, $\gamma_\varphi^{-1}=25\,\,{\rm\mu s}$, $\tau/2\pi=20$~MHz, and $\omega_\sigma/2\pi=\omega_a/2\pi=\omega_b/2\pi=\omega_d/2\pi=5$~GHz.}
\label{fig4}
\end{figure}

When the intrinsic dissipations from all constituents of the hybrid system are included, the complete master equation can be written as
\begin{equation}\label{master-magnon}
\dot{\rho}=-i[H_s^{(d)},\rho]+\tau \mathcal{L}[o]\rho
             +\sum_{\alpha=\sigma,a,b}\gamma_{\alpha}\mathcal{L}[\alpha]\rho
             +\gamma_\varphi (\sigma_z \rho\sigma_z- \rho),
\end{equation}
with $\sigma_z=|e\rangle \langle e|-|g\rangle \langle g|$, where $\gamma_\sigma$ ($\gamma_\varphi$) is the relaxation rate (pure dephasing rate) of the qubit, and $\gamma_{a}$ and $\gamma_{b}$ are the relaxation rates for the resonator and the magnon mode in the YIG sphere, respectively. By numerically solving the master equation in Eq.~(\ref{master-magnon}), we plot the maximal fidelity $\mathcal{F}_{\rm max}$ and the corresponding $t_{\rm max}$ of the prototype $W$ state $|W_3^{(1)}\rangle$ versus the Rabi frequency $\Omega_d/\tau$ in Figs.~\ref{fig4}(a) and \ref{fig4}(b). Different from the ideal case without the intrinsic dissipations of three subsystems [cf. Figs.~\ref{fig4}(a) and \ref{fig3}(b)], there is a specific Rabi frequency $\Omega_d^{\rm (opt)}$ (with $\Omega_d^{\rm (opt)}/\tau=0.0221$), where the optimum fidelity $\mathcal{F}_{\rm max}^{\rm (opt)}$ (with $\mathcal{F}_{\rm max}^{\rm (opt)}=0.936$) is reached, and the corresponding time $t_{\rm max}$ is denoted as $t_{\rm max}^{\rm (opt)}$ (with $t_{\rm max}^{\rm (opt)}=0.976\,\,{\rm\mu s}$), cf. Fig.~\ref{fig4}(b). The underlying physics of this difference is that when $\Omega_d<\Omega_d^{\rm (opt)}$, the intrinsic dissipation rather than the drive pulse dominates the dynamical behaviors of the hybrid system, which makes the state of the system dissipate to the ground state. In the weak-drive limit $\Omega_d \rightarrow 0$, the maximal fidelity $\mathcal{F}_{\rm max} \rightarrow 0$. To produce the long-lived target state $|W_3^{(1)}\rangle$ with fidelity $\mathcal{F}=\mathcal{F}_{\rm max}^{\rm (opt)}$, the parameters of the drive pulse should be set to be $\Omega_d=\Omega_d^{\rm (opt)}$ and $t_0=t_{\rm max}^{\rm (opt)}$. With these optimum parameters $\Omega_d=\Omega_d^{\rm (opt)}$ and $t_0=t_{\rm max}^{\rm (opt)}$, we show the time evolution of the fidelity $\mathcal{F}$ of $|W_3^{(1)}\rangle$ for different relaxation times $\gamma_b^{-1}$ of the magnon mode in Fig.~\ref{fig4}(c). For $\gamma_b^{-1}=1\,\,{\rm\mu s}$ ($5\,\,{\rm\mu s}$, $20\,\,{\rm\mu s}$), the fidelity $\mathcal{F}$ increases monotonically with time $t$ when $t<0.478\,\,{\rm\mu s}$ ($0.976\,\,{\rm\mu s}$, $1.572\,\,{\rm\mu s}$) and then reaches the maximum value $\mathcal{F}_{\rm max}^{\rm (opt)}=0.882$ (0.936, 0.955) at time $t=0.478\,\,{\rm\mu s}$ ($0.976\,\,{\rm\mu s}$, $1.572\,\,{\rm\mu s}$). Clearly, for a longer lifetime $\gamma_b^{-1}$ of magnons, the corresponding fidelity $\mathcal{F}_{\rm max}^{\rm (opt)}$ is higher, while the time $t_0=t_{\rm max}^{\rm (opt)}$ (for reaching $\mathcal{F}=\mathcal{F}_{\rm max}^{\rm (opt)}$) is also longer. Due to the intrinsic dissipations, the fidelity $\mathcal{F}$ decreases monotonically for $t>t_0$, and the lifetime of the target state $|W_3^{(1)}\rangle$ is limited by the lifetimes of the qubit, the resonator and the magnon mode. In addition, the other types of $W$ states with the fidelity  $\mathcal{F}=\mathcal{F}_{\rm max}^{\rm (opt)}$ , such as $|W_3^{(2)}\rangle$, $|W_3^{(3)}\rangle$ and $|W_3^{(4)}\rangle$, can also be generated via driving the qubit by appropriate pulses [cf.~Fig.~\ref{fig4}(d)].

\begin{figure}[tb]
\includegraphics[width=0.48\textwidth]{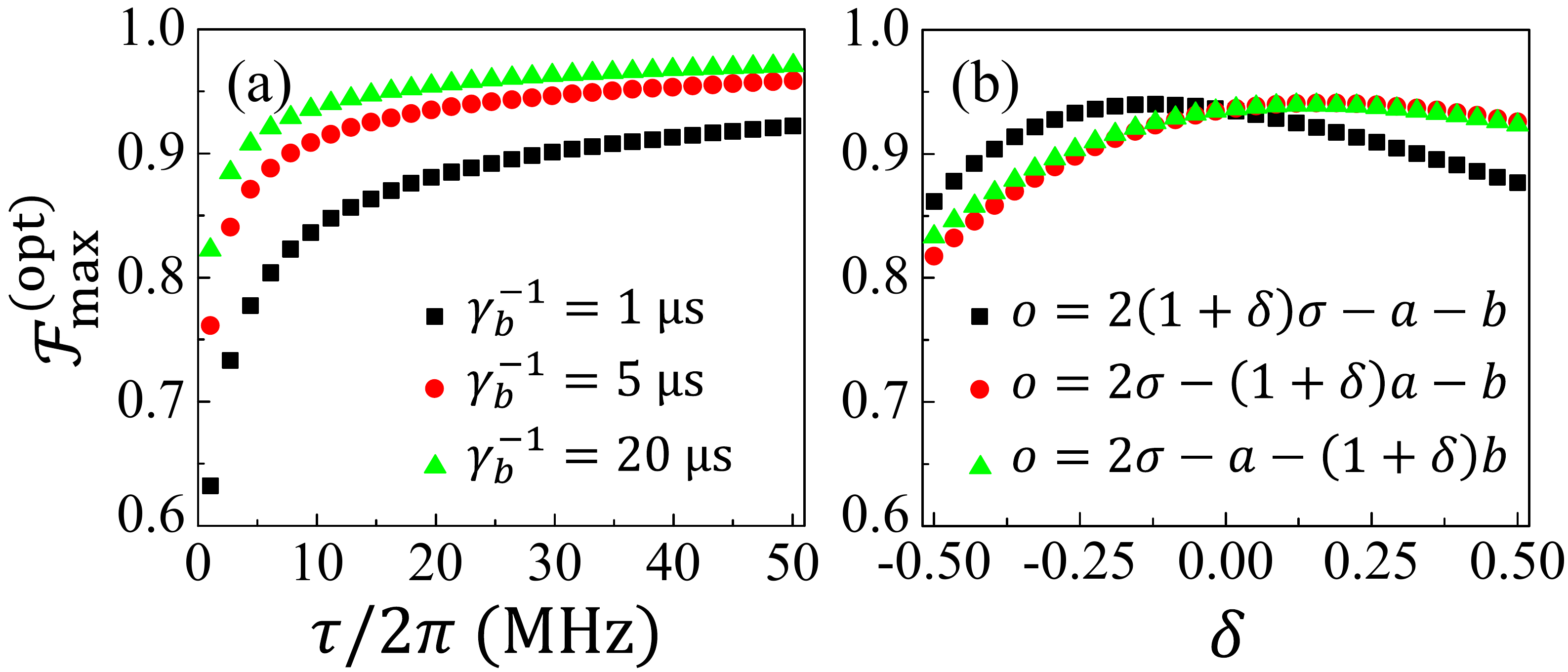}
\caption{(a) The optimum fidelity $\mathcal{F}_{\rm max}^{\rm(opt)}$ of the prototype $W$ state $|W_3^{(1)}\rangle=(|e00\rangle+|g10\rangle+|g01\rangle)/\sqrt{3}$ versus the cooperative decay rate $\tau/2\pi$ for different relaxation times of the magnon mode, $\gamma_b^{-1}=1\,\,{\rm\mu s}$ (black square), $\gamma_b^{-1}=5\,\,{\rm\mu s}$ (red circle), and $\gamma_b^{-1}=20\,\,{\rm\mu s}$ (green triangle). (b) The optimum fidelity $\mathcal{F}_{\rm max}^{\rm(opt)}$ of the prototype $W$ state $|W_3^{(1)}\rangle$ versus the deviation $\delta$ from different subsystems with $\tau/2\pi=20$~MHz and $\gamma_b^{-1}=5\,\,{\rm\mu s}$. Here, $o=2(1+\delta)\sigma-a-b$ (black square), $o=2\sigma-(1+\delta)a-b$ (red circle), and $o=2\sigma-a-(1+\delta)b$ (green triangle). In (a,b), each data point is calculated using the master equation in Eq.~(\ref{master-magnon}), where $\Omega_d=\Omega_d^{\rm(opt)}$ and $t_0=t_{\rm max}^{\rm (opt)}$. Other parameters are the same as in Fig.~\ref{fig4}.}
\label{fig5}
\end{figure}

Furthermore, we find that the optimum fidelity $\mathcal{F}_{\rm max}^{\rm (opt)}$ can be also improved by increasing the cooperative decay rate $\tau$. As shown in Fig.~\ref{fig5}(a), the optimum fidelity $\mathcal{F}_{\rm max}^{\rm(opt)}$ of the prototype $W$ state $|W_3^{(1)}\rangle$ versus the cooperative decay rate $\tau/2\pi$  increases monotonically for different relaxation times $\gamma_b^{-1}$ of the magnon mode. In the case of $\gamma_b^{-1}=1\,\,{\rm\mu s}$ ($5\,\,{\rm\mu s}$, $20\,\,{\rm\mu s}$), $\mathcal{F}_{\rm max}^{\rm(opt)}$ is improved from $0.84$ to $0.922$ ($0.911$ to $0.959$, $0.937$ to $0.971$) if the cooperative decay rate $\tau/2\pi$ increases from $10$~MHz to $50$~MHz. Experimentally, the engineered jump operator $o$ for generating the target $W$ state may be not ideal because the dissipative rates of the qubit, the resonator and the mangon mode, induced by the waveguide, are not tunable. Thus, it is necessary to investigate the effects of the deviations of the jump operator $o$ from the ideal form. In Fig.~\ref{fig5}(b), we plot the optimum fidelity $\mathcal{F}_{\rm max}^{\rm(opt)}$ of the prototype $W$ state $|W_3^{(1)}\rangle$ versus the deviation $\delta$, where $o=2(1+\delta)\sigma-a-b$ is for the deviation from the qubit (black square), $o=2\sigma-(1+\delta)a-b$ is for the deviation from the resonator (red circle), and $o=2\sigma-a-(1+\delta)b$ is for the deviation from the magnon mode (green triangle). Around the ideal value $\delta=0$, the optimum fidelity $\mathcal{F}_{\rm max}^{\rm(opt)}$ versus $\delta$ changes slowly. This robustness of $\mathcal{F}_{\rm max}^{\rm(opt)}$ against $\delta$ indicates that even if the jump operator $o$ deviates from the ideal form, but our scheme can also work well.

In the experiment, the typical frequencies for the transmon qubit, the resonator and the magnon mode can be made as $1-10$~GHz~\cite{Gu17,Yuan22}. In the numerical simulations, we set $\omega_\sigma/2\pi=\omega_a/2\pi=\omega_b/2\pi=\omega_d/2\pi=5$~GHz, which can be easily reached, because the frequencies of the transmon qubit and the magnon mode are readily tunable by controlling the bias magnetic fields. With the state-of-the-art technologies, the relaxation time $\gamma_\sigma^{-1}$ ($\gamma_a^{-1}$) of the transmon qubit (the resonator) and the pure dephasing time $\gamma_\varphi^{-1}$ of the transmon qubit can be made to be on the order of $10-100\,\,{\rm \mu s}$~\cite{Gu17}. In cavity magnonics, the typical relaxation time $\gamma_b^{-1}$ of the magnon mode is on the order of $1\,\,{\rm \mu s}$~\cite{Zhang14,Tabuchi14}, which can possibly go up to several microseconds~\cite{Spencer61} and even beyond~\cite{Jantz75} by improving the YIG sphere quality,  e.g., in surface roughness and densities of impurities and defects. Moreover, the decay rate $|\eta_\sigma|^2 \tau/2\pi=99.5$~MHz of the transmon qubit~\cite{Mirhosseini19,Hoi11} and the decay rate $|\eta_a|^2 \tau/2\pi=169$~MHz of the resonator~\cite{Haeberlein15} caused by the coplanar transmission-line waveguide were reported in experiments. Very recently, the dissipative coupling between two magnon modes in two 0.25-mm-diameter YIG spheres was engineered via a common waveguide, where the decay rate $|\eta_b|^2 \tau/2\pi$ of each magnon mode induced by the waveguide is $8.5$~MHz~\cite{Li22}. If the diameter of the YIG sphere is increased to 0.4 mm (0.6 mm), the decay rate of each magnon mode due to the waveguide will be $|\eta_b|^2 \tau/2\pi=34.8$~MHz (117.5~MHz), because the decay rate of magnons due to the waveguide is proportional to the volume of the YIG sphere~\cite{Tabuchi14}. Therefore, our scheme is experimentally feasible with the currently available parameters.

\section{Discussions and conclusions}\label{conclusions}

Usually, the common reservoir is artificial and can only induce the cooperative relaxation in dissipatively coupled systems~\cite{Gu17,Wang20-1,Wang22,Zhang19-Baranger,Zhan22,Zou21}. In our scheme (cf.~Sec.~\ref{qubit-photon-magnon}), the common reservoir is an open waveguide, and no cooperative pure dephasing of the system is induced (see the Appendix~\ref{Appendix-A}). If the qubit and the two resonators are coupled via a common pure dephasing reservoir, the corresponding master equation of the system can be written as Eq.~(\ref{master}) with the jump operator $o=\eta_\sigma \sigma+\eta_a a+\eta_b b$ replaced by $O=\eta_\sigma \sigma^\dag\sigma+\eta_a a^\dag a+\eta_b b^\dag b$. Different from the cooperative relaxation, the cooperative pure dephasing cannot induce the energy exchange among the three quantum subsystems. Thus, the proposed scheme does not work for a common pure dephasing reservoir.

In addition, it should be emphasized that the common reservoir (i.e., the cooperative dissipation) plays a crucial role in our scheme.
Due to the drive pulse on the qubit, the ternary system evolves from $|g00\rangle$ to $|e00\rangle$. Then, the component of the bright state $|B\rangle$ (i.e., the nontarget $W$ state) in the state $|e00\rangle$ decays to the vacuum state $|g00\rangle$ via radiating the energy into the common reservoir, while the component of the dark state $|D\rangle$ (i.e., the target $W$ state) in the state $|e00\rangle$ is steady because it decouples from the common reservoir [cf.~Eqs.~(\ref{e00}) and (\ref{dark}) and related discussions]. Without the common reservoir, the three eigenvalues of the closed system in the one-excitation subspace will be real, which means that the corresponding three eigenvectors are all dark states because their decay rates are zero. As a result, both components of target $W$ state and nontarget $W$ states in the state $|e00\rangle$ are dark states. Thus, the target $W$ state cannot be generated in the closed system.

In summary, we presented a scheme to generate tripartite $W$ states in a ternary system consisting of one qubit and two resonators, which are dissipatively coupled via a common reservoir. With appropriate values of the system parameters, the $W$ state can be prepared by pumping the qubit with a drive pulse. This scheme is easy to implement in experiments since neither performing measurements nor adjusting the system parameters is required. In addition, because the generated $W$ state is a dark state of the system, it is steady and has a very long lifetime. To show the validity of the scheme, we apply our scheme in a hybrid qubit-photon-magnon system. Due to the intrinsic dissipations from all subsystems, the lifetime of the generated qubit-photon-magnon $W$ state is determined by the lifetimes of the qubit, the resonator and the magnon mode. Besides magnon based hybrid systems, our scheme can be also applicable to waveguide QED systems~\cite{Gu17,Zhang19-Baranger,Zhan22}, dissipatively coupled spins mediated by a magnetic environment~\cite{Zou21}, dissipatively coupled qubits through plasmons~\cite{Gonzalez-Tudela11,Martin-Cano11}, and so on.

\section*{Acknowledgments}

The numerical simulations of the master equation were performed using QuTiP~\cite{Johansson12,Johansson13}. This work is supported by the National Natural Science Foundation of China (Grants No.~12205069, No.~12204139 and No. U21A20436), the Key-Area Research and Development Program of GuangDong province (Grant No.~2018B030326001) and the key program of the Natural Science Foundation of Anhui (Grant No. KJ2021A1301).

\appendix

\section{Derivation of the master equation in Eq.~(\ref{master})}\label{Appendix-A}

As shown in Fig.~\ref{fig1}(a), the qubit and the two resonators simultaneously interact with a common reservoir.
We assume that the reservoir consists of many bosonic modes, where the $k$th mode is described by the annihilation (and creation) operator $c_k$ (and $c_k^\dag$) and frequency $\omega_k$. The quantum dynamics of the ternary system and the reservoir is governed by the following Hamiltonian:
\begin{eqnarray}\label{A1}
H_{\rm tot}&=&H_0 + H_{\rm int},\nonumber\\
H_0&=&H_s+\sum_k \omega_k c_k^\dag c_k,\nonumber\\
H_{\rm int}&=&\sum_k\sum_{\alpha=\sigma,a,b} \lambda_{k\alpha} (c_k^\dag\alpha e^{-i\phi_{k\alpha}}+ c_k\alpha^\dag  e^{i\phi_{k\alpha}}),
\end{eqnarray}
with the Hamiltonian $H_s$ of the ternary system given in Eq.~(\ref{Hamiltonian}). $H_0$ includes the free energy of the ternary system and the free energy of the reservoir, and $H_{\rm int}$ denotes the interaction between the ternary system and the reservoir, where $\lambda_{k\alpha}$ is the coupling strength of the subsystem $\alpha$ ($\alpha=\sigma,a,b$) to the $k$th mode of the reservoir, $\phi_{k\alpha}=(\omega_k/\upsilon)x_\alpha$ is the phase delay of the $k$th mode at the location $x_\alpha$, and $\upsilon$ is the speed of light.

When the qubit and the two resonators are resonant, i.e., $\omega_\sigma=\omega_a=\omega_b=\omega_0$, in the interaction picture, the Hamiltonian $\mathcal{V}(t)=e^{iH_0 t}H_{\rm int}e^{-iH_0 t}$ is given by
\begin{eqnarray}\label{A2}
\mathcal{V}(t)&=&\sum_k \lambda_{k}
                \left[c_k^\dag o e^{-i(\omega_0-\omega_k)t}
                 + c_k o^\dag  e^{i(\omega_0-\omega_k)t}\right].
\end{eqnarray}
Here we defined a collective jump operator
\begin{eqnarray}\label{A3}
o=\eta_\sigma \sigma+\eta_a a+\eta_b b
\end{eqnarray}
with $\eta_\alpha=(\lambda_{k\alpha}/\lambda_{k})e^{-i\phi_{k\alpha}}$ and introduced an effective coupling strength $\lambda_{k}$ between the ternary system and the $k$th reservoir mode. Taking a trace over the reservoir coordinates under the Born-Markov approximation, the density $\tilde{\rho}(t)$ of the ternary system satisfies~\cite{Scully97}
\begin{eqnarray}\label{A4}
\dot{\tilde{\rho}}(t)&=&-i\,\rm{Tr}_{\rm r}[\mathcal{V}(t),\tilde{\rho}(0) \otimes \rho_r(0)]\nonumber\\
             & &-\rm{Tr}_{\rm r}\int_{0}^{t}dt'[\mathcal{V}(t),[\mathcal{V}(t'),\tilde{\rho}(t) \otimes \rho_r(0)]],
\end{eqnarray}
where $\rho_r(0)$ is the density matrix of the reservoir at $t=0$. In our paper, the considered reservoir is at zero temperature and $\rho_r(0)$ is the multi-mode extension of the thermal operator. It can be easily shown that
\begin{eqnarray}\label{A5}
\langle c_k\rangle &=&\langle c_k^\dag\rangle=0,\nonumber\\
\langle c_{k}^\dag c_{k'}\rangle&=&0,\nonumber\\
\langle c_k c_{k'}^\dag\rangle&=&\delta_{kk'},\nonumber\\
\langle c_{k} c_{k'}\rangle&=&\langle c_{k}^\dag c_{k'}^\dag\rangle=0.
\end{eqnarray}
Here $\langle \mathcal{O}\rangle \equiv {\rm Tr}[\mathcal{O}\rho_r(0)]$ is the expectation value of any reservoir operator $\mathcal{O}$. With the relations in Eq.~(\ref{A5}), we insert the Hamiltonian $\mathcal{V}(t)$ in Eq.~(\ref{A2}) into the equation (\ref{A4}) of motion and obtain
\begin{eqnarray}\label{A6}
\dot{\tilde{\rho}}(t)&=&\int_{0}^{t}dt'\sum_k \lambda_k^2e^{i(\omega_0-\omega_k)t'}[o\tilde{\rho}(t) o^\dag
                  - o^\dag o\tilde{\rho}(t)]+{\rm H.c.}\,.~~~~~~
\end{eqnarray}
Using the standard identity
\begin{eqnarray}\label{A7}
\lim_{t\rightarrow +\infty}\int_{0}^{t}dt'e^{i(\omega_0-\omega_k)t'}=\pi \delta(\omega_0-\omega_k),
\end{eqnarray}
the master equation in Eq.~(\ref{A6}) becomes
\begin{eqnarray}\label{A8}
\dot{\tilde{\rho}}(t)&=&\tau[2o\tilde{\rho}(t) o^\dag- o^\dag o\tilde{\rho}(t)-\tilde{\rho}(t)o^\dag o],
                         ~~~
\end{eqnarray}
with the decay rate $\tau=\sum_k \pi \lambda_k^2\delta(\omega_0-\omega_k)$ of the ternary system.

Usually, the main contribution to the decay rate $\tau$ arises from the modes of the reservoir with frequency $\omega_k \approx \omega_0$. Because the phase $\phi_{k\alpha}=(\omega_k/\upsilon)x_\alpha$ in the jump operator $o$ varies little around $\omega_k=\omega_0$, we can take the approximation $\phi_{k\alpha}=\phi_{\alpha}\approx (\omega_0/\upsilon)x_\alpha$. It should be pointed out that $\tilde{\rho}(t)$ is the density matrix of the system in the interaction picture, which is related to the density matrix $\rho$ of the system in the Schr\"{o}dinger picture via the relation $\rho=e^{-iH_s t}\tilde{\rho}(t)e^{iH_s t}$. It follows from Eq.~(\ref{A8}) that the density matrix $\rho$ satisfies
\begin{eqnarray}\label{A9}
\dot{\rho}&=&-i[H_s,\rho]+\tau(2o\rho o^\dag- o^\dag o\rho-\rho o^\dag o).
\end{eqnarray}
This is just Eq.~(\ref{master}) in the main text.

It should be noted that the master equation in Eq.~(\ref{A9}) is valid only when the system parameters satisfy the following three conditions. (i) The qubit and the two resonators are nearly resonant, i.e., $\omega_\sigma \approx \omega_a \approx \omega_b\approx \omega_0$. This condition was used in the derivation of Eq.~(\ref{A2}). (ii) The value of $\{|\eta_\sigma|^2\tau,|\eta_a|^2\tau,|\eta_b|^2\tau\}_{\rm max}$ cannot be too large. In Eq.~(\ref{A1}), the Hamiltonian $H_{\rm int}$ is obtained under the rotating-wave approximation by neglecting the fast-oscillating terms, and the Born-Markov approximation is used in the derivation of Eq.~(\ref{A4})~\cite{Scully97}. Both the approximations require that the coupling between the system and the common reservoir cannot be too strong (related to the values of $\{|\eta_\sigma|^2\tau,|\eta_a|^2\tau,|\eta_b|^2\tau\}$). In Ref.~\cite{Wang19}, the experimental results show that both the rotating-wave approximation and the Born-Markov approximation are still valid, even if the decay rate $|\eta_b|^2\tau/2\pi$ of the magnon mode induced by the common reservoir is up to $880$~MHz, where the frequency of the magnon mode is $\omega_b/2\pi \approx 5$~GHz. (iii) The drive pulse on the qubit should be weak (i.e., $\Omega_d\ll \tau$). In our scheme, generating the $W$ state requires a drive pulse on the qubit (cf.~Secs.~\ref{Generating-W-states} and \ref{qubit-photon-magnon}). However, the effect of the drive pulse on the dissipative term in the master equation (\ref{A9}) is neglected, which is reasonable for a weak drive pulse. In quantum optics, this approximation is widely used~\cite{Scully97}. In the numerical simulations of this paper, the three parameter conditions in (i) to (iii) are safely satisfied (see the main text).

\end{document}